\documentclass[aps,pra,twocolumn,showpacs,superscriptaddress,floatfix]{revtex4}
\usepackage{graphicx}
\usepackage{bm}
\begin{document}
\title{Reactions of ultracold alkali metal dimers}
\author{Piotr S. \.Zuchowski}
\email{E-mail: Piotr.Zuchowski@durham.ac.uk}%
\affiliation{Department of Chemistry, Durham University, South
Road, Durham, DH1~3LE, United Kingdom}
\author{Jeremy M. Hutson}
\email{E-mail: J.M.Hutson@durham.ac.uk}%
\affiliation{Department of Chemistry, Durham University, South Road,
Durham, DH1~3LE, United Kingdom}

\date{\today}

\begin{abstract}
We investigate the energetics of reactions involving pairs of
alkali metal dimers. Atom exchange reactions to form
homonuclear dimers are energetically allowed for some but not
all of the heteronuclear dimers. We carry out high-level
electronic structure calculations on the potential energy
surfaces of all the heteronuclear alkali metal trimers and show
that trimer formation reactions are always energetically
forbidden for low-lying singlet states of the dimers. The
results have important implications for the stability of
quantum gases of alkali metal dimers.
\end{abstract}


\maketitle

It has recently become possible to create samples of alkali
metal dimers in deeply bound states at temperatures below
$10^{-6}$~K \cite{Sage:2005, Ni:KRb:2008, Lang:ground:2008,
Danzl:v73:2008, Viteau:2008, Deiglmayr:2008, Mark:2009,
Haimberger:2009, Danzl:ground:2010}. For KRb \cite{Ni:KRb:2008}
and Cs$_2$ \cite{Danzl:ground:2010}, molecules are first formed
in high-lying vibrational states by magnetoassociation and then
coherently transferred to the absolute ground state by
stimulated Raman adiabatic passage (STIRAP). These capabilities
open up new possibilities for ultracold chemistry, for creating
strongly interacting quantum gases, and for producing tunable
models of important condensed-phase systems
\cite{Carr:NJPintro:2009}.

Ultracold molecules are usually confined in very shallow traps.
Any collision that transfers internal energy into relative
kinetic energy is likely to eject both collision partners from
the trap. If both species are in their absolute ground state,
inelastic collisions are impossible, but there remains the
possibility of {\em reactive} collisions. Indeed, for fermionic
$^{40}$K$^{87}$Rb, Ospelkaus {\em et al.}\
\cite{Ospelkaus:react:2010} have carried out detailed studies
of the exothermic atom exchange reaction,
\begin{equation}
\hbox{KRb} + \hbox{KRb} \longrightarrow \hbox{K}_2
+ \hbox{Rb}_2. \label{eq:exch}
\end{equation}
When all the molecules are in the same nuclear spin state, the
reaction rate is strongly suppressed by the Pauli principle.
However, if some of the molecules are transferred into a
different spin state, the reaction proceeds very fast and the
molecules are lost from the trap.

As will be seen below, atom exchange reactions analogous to
(\ref{eq:exch}) are energetically allowed for some alkali metal
dimers but forbidden for others. However, even when atom
exchange is forbidden, there remains in principle the
possibility of atom transfer reactions such as
\begin{equation}
\hbox{KRb} + \hbox{KRb} \longrightarrow \hbox{K}
+ \hbox{KRb}_2 \hbox{\quad or \quad K}_2\hbox{Rb} +
\hbox{Rb}. \label{eq:trans}
\end{equation}
In a simple pairwise-additive model of the energetics, the
reactants in such a reaction have two nearest-neighbor
interactions and the products have three, so some researchers
have anticipated that the trimer formation reactions would be
energetically allowed. However, pairwise-additive models are
known to be very poor for the quartet excited states of alkali
metal trimers \cite{Soldan:2003} and are likely to be even
poorer for the doublet ground states. The principle purpose of
this paper is to explore the energetics of trimer formation
reactions such as (\ref{eq:trans}). We will demonstrate that,
for singlet alkali metal dimers in levels near the potential
minimum, trimer formation reactions are in fact always
energetically forbidden.

Before proceeding to the trimer formation reactions, we briefly
consider atom exchange reactions analogous to (\ref{eq:exch})
for the heteronuclear dimers formed from the alkali metals Li,
Na, K, Rb and Cs. All the homonuclear and heteronuclear dimers
except LiRb have been studied in detail by high-resolution
spectroscopy, and dissociation energies $D_e$ accurate to $\pm
1$ cm$^{-1}$ or better have been extracted as listed in Table
\ref{tab:diss}. The energy changes for the atom exchange
reactions can therefore be calculated directly from experiment,
and are summarized in Table \ref{tab:exch}. The values given
are taken from dissociation energies $D_e$ measured to the
dimer equilibrium geometries and so are subject to small
corrections for the differences in zero-point energy between
reactants and products. These corrections can be up to +25
cm$^{-1}$ for LiX systems but are less than $\pm2$ cm$^{-1}$
for the remainder. It may therefore be concluded that all the
heteronuclear Li dimers and KRb will be subject to reactive
trap loss, but all the remainder should be stable with respect
to atom exchange collisions in their ground rovibronic state.
\begin{table*}
\caption{Dissociation energies $D_e$ (in cm$^{-1}$) for alkali
metal dimers. The quantities in parentheses are uncertainties
in the final digit(s). \label{tab:diss}}
\begin{ruledtabular}
\begin{tabular}{lccccc}
    &  Li       &  Na          &   K      & Rb      & Cs \\
Li  & 8516.768(8)\footnote{Ref.\ \onlinecite{Coxon:2006}}
    & 7105.5(1.0)\footnote{Ref.\ \onlinecite{Fellows:1991}}
    & 6216.886(100)\footnote{Ref.\ \onlinecite{Tiemann:2009}}
    & 5946(100)\footnote{The binding energy for LiRb is not
    available from experiment, so this value is calculated
    using the AQCC method described in this paper.}
    & 5875.542(5)\footnote{Ref.\ \onlinecite{Grochola:2009}}\\
Na && 6022.0286(53)\footnote{Ref.\ \onlinecite{Jones:1996}}
    & 5273.62(10)\footnote{Ref.\ \onlinecite{Gerdes:2008}}
    & 5030.502(10)\footnote{Ref.\ \onlinecite{Pashov:NaRb:2005}}
    & 4954.237(100)\footnote{Ref.\ \onlinecite{Docenko:2006}}\\
K &&& 4450.906(50)\footnote{Ref.\ \onlinecite{Pashov:2008}}
    & 4217.815(10)\footnote{Ref.\ \onlinecite{Pashov:2007}}
    & 4069.208(40)\footnote{Ref.\ \onlinecite{Ferber:2009}}\\
Rb &&&& 3993.47(18)\footnote{Ref.\ \onlinecite{Seto:2000}}
    & 3836.14(50)\footnote{Ref.\ \onlinecite{Fellows:1999}}\\
Cs &&&&& 3649.695(2)\footnote{$D_0$ from Ref.\ \onlinecite{Danzl:v73:2008}
    and zero-point energy from Ref.\ \onlinecite{Amiot:2002}}\\
\end{tabular}
\end{ruledtabular}
\end{table*}

\begin{table}[tbhdp]
\caption{Energy changes $\Delta E_2$ for the reactions
2XY$\to$X$_2$+Y$_2$ (in cm$^{-1}$). The quantities in
parentheses are uncertainties in the final digit(s).
\label{tab:exch}}
\begin{ruledtabular}
\begin{tabular}{lccccc}
   & Na          &   K         & Rb          & Cs \\ \hline
Li &  $-328(2)$  & $-533.9(3)$ & $-618(200)$ & $-415.38(2)$ \\
Na   &           &  74.3(3)    & 45.5(5)     & 236.75(20)   \\
K    &           &             & $-8.7(9)$   & 37.81(13)    \\
Rb   &           &             &             & 29.1(1.5)    \\
\end{tabular}
\end{ruledtabular}
\end{table}

Trimer formation reactions cannot be considered in a similar
way because an experimental binding energy is available only
for Li$_3$ \cite{Wu:1976} and not for any of the heteronuclear
trimers. We have therefore carried out electronic structure
calculations for all the homonuclear and heteronuclear alkali
metal trimers, using the multireference average-quadratic
coupled-cluster method (AQCC). All calculations used the MOLPRO
package \cite{MOLPRO_brief:2006}. The alkali atoms were
described in a single-electron model and the core-valence
interaction was taken into account using an effective core
potential (ECP) with a core polarization potential (CPP). We
used the ECPxSDF family of core potentials, developed by the
Stuttgart group \cite{Fuentealba:CPP1:1982,
Fuentealba:CPP1:1983}, with core polarization potentials based
on those of M\"uller and Meyer \cite{Muller:CPP2:1984}. We
obtained modified values of the M\"uller-Meyer cutoff parameter
(0.95 for Li, 0.82 for Na, 0.36 for K, 0.265 for Rb and 0.24
for Cs) that reproduce the experimental bond lengths of the
ground-state homonuclear alkali dimers at the same level of
theory. We used the uncontracted $sp$ basis sets designed for
ECPxSDF core potentials \cite{Fuentealba:CPP1:1982,
Fuentealba:CPP1:1983}, augmented by additional $s$, $p$, $d$
and $f$ functions \cite{alkali-basis:CPP:2010}. With these
polarization potentials and basis sets we reproduced the
singlet binding energies for homonuclear alkali metal dimers
with an accuracy better than 1\% for Li$_2$, Na$_2$, Rb$_2$ and
2\% for K$_2$ and Cs$_2$. The binding energies for the
heteronuclear dimers are as good as for the homonuclear dimers,
except for LiCs, for which the error in the binding energy was
+2.5\%.

\begin{table}[tbhdp]
\caption{Atomization energies and equilibrium geometries of the
X$_2$Y trimers from AQCC calculations, together with energy
changes $\Delta E_3$ for the reactions 2XY$\to$X$_2$Y + Y,
obtained by combining the trimer results with dimer
dissociation energies obtained with the same method.
 \label{tab:trans}}
\begin{ruledtabular}
\begin{tabular}{llccccc}
\multicolumn{5}{l}{Atomization energy (cm$^{-1}$)}
                               & X    &      \\ \hline
  &    & Li      & Na     &   K       & Rb      & Cs    \\
  & Li &  13189  & 9977   & 8341     & 7982     & 8378  \\
  & Na &  11583  & 8113   & 7125     & 6783     & 7140  \\
Y & K  &  10681  & 7795   & 6258     & 5902     & 5890  \\
  & Rb &  10499  & 7649   & 6080     & 5685     & 5661  \\
  & Cs &  11073  & 8128   & 6211     & 5781     & 5494  \\
\end{tabular}
\end{ruledtabular}
\smallskip
\begin{ruledtabular}
\begin{tabular}{llccccc}
\multicolumn{4}{l}{$r_{\rm X_1Y}$,$r_{\rm X_2Y}$,$r_{\rm XX}$ (\AA)}
                           &   X      &          &  \\ \hline
  &    & Li           & Na          &   K           & Rb            & Cs          \\
  & Li & 2.8,2.8,3.2  & 3.0,3.0,4.0 &  3.5,3.5,5.3  & 3.6,3.6,5.9   & 4.1,4.1,4.7 \\
  & Na & 3.1,3.5,2.7  & 3.2,3.2,4.2 &  3.7,4.4,4.0  & 4.0,4.4,4.2   & 4.1,4.4,4.6 \\
Y & K  & 3.5,4.3,2.8  & 3.7,3.7,3.9 &  4.1,4.1,5.2  & 4.2,4.2,5.7   & 4.4,5.5,4.8 \\
  & Rb & 3.6,4.5,2.8  & 3.8,3.8,3.8 &  4.2,5.3,4.1  & 4.4,4.4,5.5   & 4.6,5.5,4.7 \\
  & Cs & 3.8,3.8,3.1  & 4.0,4.0,3.7 &  4.5,4.5,4.9  & 4.6,4.6,5.5   & 4.8,4.8,5.7 \\
\end{tabular}
\end{ruledtabular}
\smallskip
\begin{ruledtabular}
\begin{tabular}{llccccc}
\multicolumn{4}{l}{$\Delta E_3$ (cm$^{-1}$)}
                        &   X  &      &      \\ \hline
  &    & Li     & Na    &   K   & Rb      & Cs       \\
  & Li & 3759   & 4145  &  3979  & 3910   & 3660     \\
  & Na & 2539   & 3843  &  3281  & 3287   & 2962     \\
Y & K  & 1639   & 2611  &  2460  & 2444   & 2264     \\
  & Rb & 1393   & 2421  &  2266  & 2295   & 2101     \\
  & Cs & 965    & 1974  &  1943  & 1981   & 1958     \\
\end{tabular}
\end{ruledtabular}
\end{table}

To understand the doublet states of heteronuclear alkali metal
trimers, it is useful first to consider the homonuclear
systems. The important molecular orbitals are those formed from
the outermost $s$ orbitals on each atom. At an equilateral
triangular configuration ($D_{3h}$ symmetry), the two
lowest-lying molecular orbitals of a homonuclear trimer have
$a_1$ and $e$ symmetry. The lowest doublet state has
configuration $a_1^2 e^1$. It is therefore orbitally
degenerate, with $^2E$ symmetry, and is subject to a
Jahn-Teller distortion to an isosceles geometry ($C_{2v}$) that
splits the $e$ orbitals into $a_1$ and $b_2$ components: the
$b_2$ orbital has a node between the two equivalent atoms. The
equilibrium structures of the homonuclear trimers all have
$C_{2v}$ geometries with ground states of $^2B_2$ symmetry.

\begin{figure}
\includegraphics[width=0.85\linewidth]{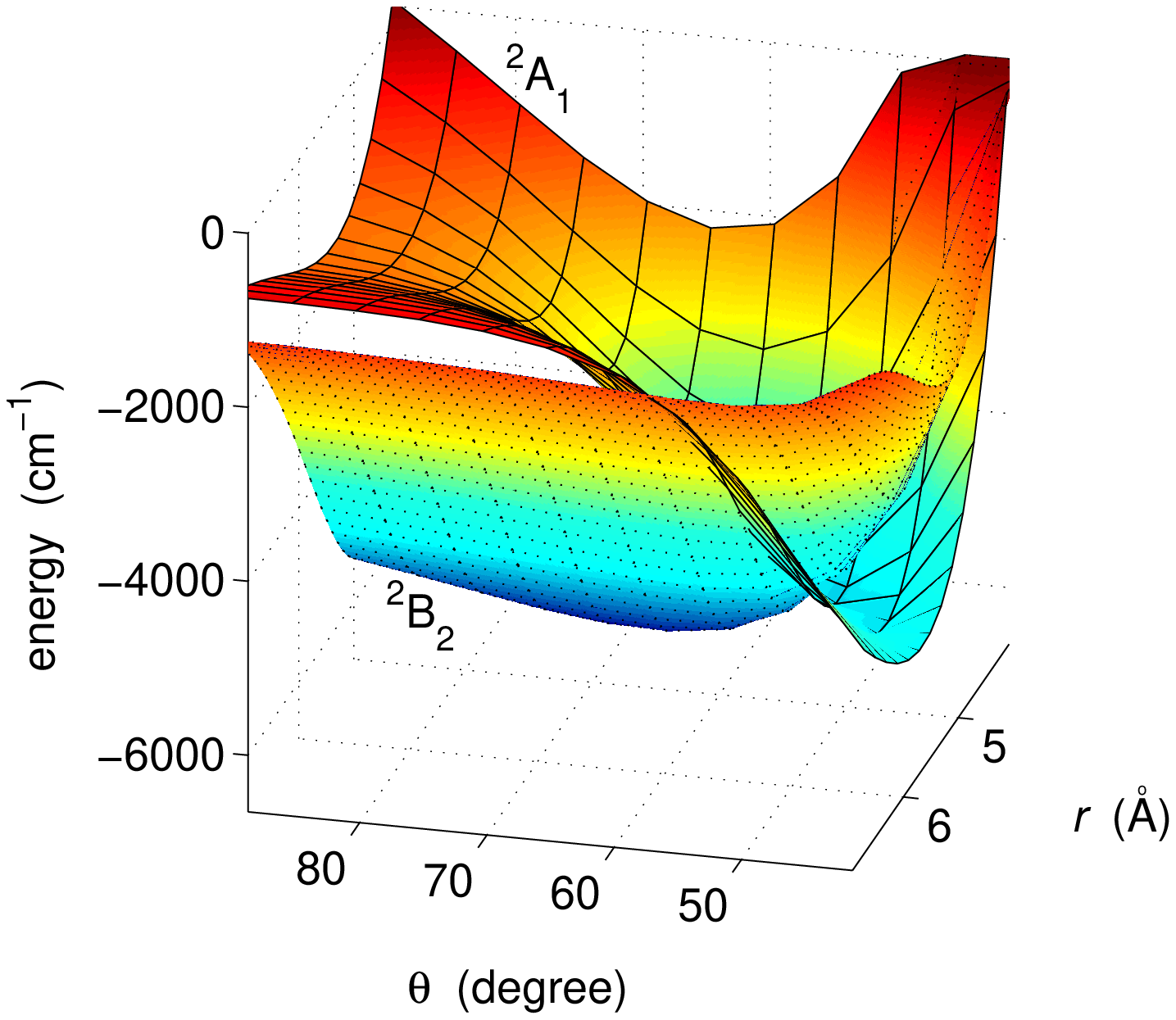}
\bigskip\bigskip
\includegraphics[width=0.95\linewidth]{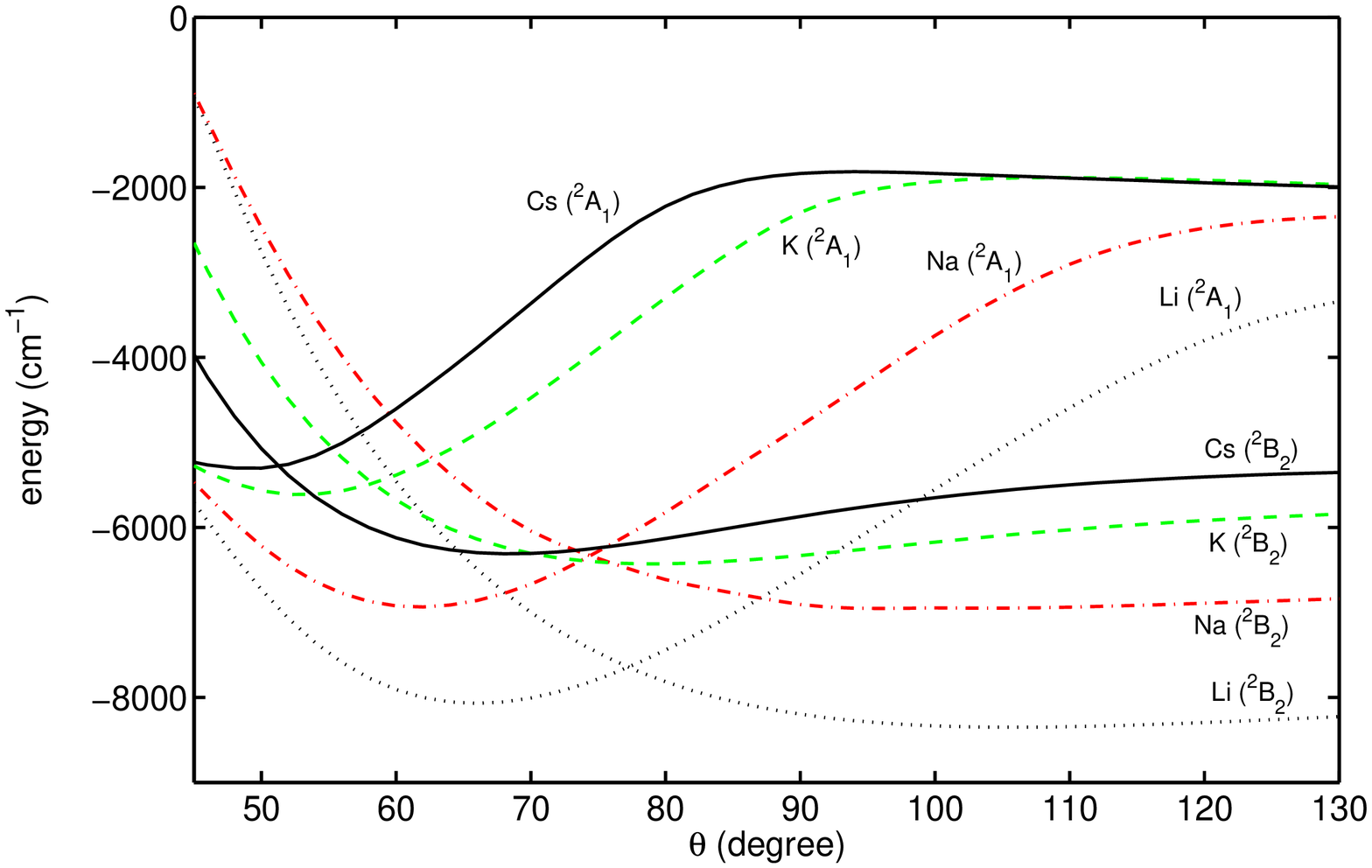}
\caption{(Color online)
Upper panel: The intersecting potential energy surfaces for
$^2A_1$ and $^2B_2$ states of Rb$_2$Cs at $C_{2v}$ geometries,
as a function of the RbCs bond length $r$ and the bond angle
$\theta$. Lower panel: curves for the $^2A_1$ and $^2B_2$
states of Rb$_2$X systems as a function of bond angle $\theta$,
with the bond length optimized at each angle. For X=Li,Cs,K the
minima for $^2B_2$ states are the absolute minima, while for
Rb$_2$Na system the absolute minimum originates from the
distorted $^2A_1$ minimum geometry.} \label{rb2cs}
\end{figure}

For a heteronuclear trimer X$_2$Y, the symmetry is always
$C_{2v}$ or $C_s$. For $C_{2v}$ geometries, the upper $a_1$
orbital and the $b_2$ orbital are close together in energy and
the minimum (restricted to $C_{2v})$ may be on either the
$^2A_1$ surface or the $^2B_2$ surface. We have therefore
calculated the energy for all the heteronuclear trimers in both
$^2A_1$ and $^2B_2$ states for $C_{2v}$ geometries. Typical
results are shown for Rb$_2$Cs in the upper panel of Fig.\
\ref{rb2cs}. The geometry is specified by a bond length
$r=r_{\rm X_1Y}=r_{\rm X_2Y}$ and the angle $\theta$ between
the two XY bonds. It may be seen that the two surfaces
intersect at an angle near $\theta=50^\circ$: since the two
states have the same symmetry at $C_s$ geometries, they
actually intersect {\em only} at $C_{2v}$ geometries, producing
a seam of conical intersections there. An alternative
representation of the results, for all the Rb$_2$X systems, is
shown in the lower panel of the Figure: in this case $r$ has
been optimized to find the energy minimum for each value of
$\theta$, producing intersecting potential curves rather than
2-dimensional surfaces. The minima on the two surfaces are
usually close together in energy (always within 1000 cm$^{-1}$,
but often within 200 cm$^{-1}$). However, the $^2B_2$ minimum
is below the $^2A_1$ minimum for all the trimers except the 7
heteronuclear X$_2$Na and Cs$_2$X species; for Rb$_2$Cs, shown
in Fig.\ \ref{rb2cs}, the $^2B_2$ minimum is near
$\theta=63^\circ$. The equilibrium geometries and energies for
both states are provided as supplementary material
\cite{trimer-EPAPS}.

For heteronuclear trimers there is the additional possibility
of distortion to a lower-symmetry $C_s$ (scalene) geometry. We
have therefore explored whether such distortions lower the
trimer energies. At $C_s$ geometries the valence orbitals
formed from atomic $s$ orbitals are all of $a'$ symmetry, so
both low-lying states have $^2A'$ symmetry and can mix.
Nevertheless, in most cases it is clear whether the
singly-occupied orbital has bonding character ($a_1$-like) or
antibonding character ($b_2$-like) between the two like atoms.
For Cs$_2$Li, where the $^2A_1$ state was already below the
$^2B_2$ state, distortion does not lower the energy and the
equilibrium geometry has $C_{2v}$ symmetry. However, for all
the other systems the geometry corresponding to the $^2A_1$
minimum is in fact a saddle point on the full 3-dimensional
surface: for Li$_2$Na, K$_2$Na, Rb$_2$Na, Cs$_2$Li, Cs$_2$Na,
Cs$_2$K and Cs$_2$Rb, this simply deepens the minimum. For
Li$_2$K, Li$_2$Rb, K$_2$Rb, K$_2$Cs, Rb$_2$Na, the distortion
produces a $^2A'$ state whose absolute minimum (of $C_s$
symmetry) is lower in energy than the $^2B_2$ state (which
always retains an equilibrium geometry of $C_{2v}$ symmetry).
However, for Li$_2$Cs, Na$_2$Li, Na$_2$K, Na$_2$Rb, Na$_2$Cs,
K$_2$Li, Rb$_2$Li, Rb$_2$K and Rb$_2$Cs the energy gained by
distortion is not enough and the $^2B_2$ state of $C_{2v}$
geometry remains the absolute minimum.

Table \ref{tab:trans} summarizes the trimer atomization
energies, equilibrium geometries and the energy change for the
trimer formation reactions for all the alkali metal trimers
from Li to Cs. It may be seen that all the trimer formation
reactions (from singlet dimers) are substantially endoergic.
Trimer formation reactions will therefore not take place for
alkali metal dimers formed in singlet states near the bottom of
the potential well. However, trimers may of course still be
formed from dimers in triplet states, which are much more
weakly bound, or from high-lying vibrational states, including
Feshbach molecules.

The trimer energies are always substantially {\em below} the
energy of any atom + diatom arrangement of the same atoms. The
entrance channels of chemical reactions between alkali metal
atoms and singlet dimers are thus likely to be barrierless, as
shown by Tscherbul {\em et al.} \cite{Tscherbul:Rb2Cs:2008} for
Rb + RbCs (though of course the reactions themselves will not
always be energetically allowed). However, a full treatment of
the dynamics of these reactions would require a detailed study
of the potential energy surfaces for at least the two
lowest-lying electronic states and the interactions between
them. This contrasts with the situation for reactions involving
spin-stretched states of alkali metal atoms and triplet dimers,
which have been studied using single electronic surfaces for
the quartet states of the trimers \cite{Soldan:2002,
Cvitas:bosefermi:2005, Cvitas:hetero:2005, Quemener:2005,
Cvitas:li3:2007, Hutson:IRPC:2007}.

Our atomization energies for homonuclear systems may be
compared with 13436 cm$^{-1}$ for Li$_3$ from multi-reference
configuration interaction (MRCI) calculations
\cite{Kramer:1999}, and 5437.1 cm$^{-1}$ for Cs$_3$ from full
configuration interaction (CI) calculations
\cite{Guerout:2009}. Our values for trimers containing Li, Na
and K also agree well (within 1000 cm$^{-1}$) with early CI work
by Pavolini and Spiegelmann \cite{Pavolini:1987}. In all cases
the calculations used effective core potentials similar to
those in the present work.

The present results for trimer energies may be rationalized
using a very simple model. In the simplest form of H\"uckel
theory, considering only one $s$ orbital on each atom, with a
bond integral $\beta$, a homonuclear dimer in a singlet state
has binding energy $2|\beta|$. An equilateral trimer has
binding energy $3|\beta|$, while a linear trimer has binding
energy $2\sqrt{2}|\beta|$. An atom transfer reaction such as
(\ref{eq:trans}) is therefore endoergic by $|\beta|$ or
slightly more, i.e. by about half the dimer binding energy.
This is quite different from the result predicted by pairwise
additivity. However, simple orbital-based models of chemical
bonding must be treated with caution for the alkali metals,
because they have low-lying $p$ orbitals that often make
important contributions to bonding. Ion-pair states can also be
important. Thus, while H\"uckel theory can be used to
rationalize the results of the present work, it could not have
been used to predict them. The use of high-level electronic
structure calculations, as in the present work, is essential to
obtain reliable conclusions.

We are grateful to Adam Miller for assistance in compiling the
experimental results on alkali dimer binding energies. This
work is supported by EPSRC under collaborative projects CoPoMol
and QuDipMol of the ESF EUROCORES Programme EuroQUAM.

\bibliography{../../all}

\begin{thebibliography}{43}
\expandafter\ifx\csname natexlab\endcsname\relax\def\natexlab#1{#1}\fi
\expandafter\ifx\csname bibnamefont\endcsname\relax
  \def\bibnamefont#1{#1}\fi
\expandafter\ifx\csname bibfnamefont\endcsname\relax
  \def\bibfnamefont#1{#1}\fi
\expandafter\ifx\csname citenamefont\endcsname\relax
  \def\citenamefont#1{#1}\fi
\expandafter\ifx\csname url\endcsname\relax
  \def\url#1{\texttt{#1}}\fi
\expandafter\ifx\csname urlprefix\endcsname\relax\def\urlprefix{URL }\fi
\providecommand{\bibinfo}[2]{#2}
\providecommand{\eprint}[2][]{\url{#2}}

\bibitem[{\citenamefont{Sage et~al.}(2005)\citenamefont{Sage, Sainis, Bergeman,
  and DeMille}}]{Sage:2005}
\bibinfo{author}{\bibfnamefont{J.~M.} \bibnamefont{Sage}},
  \bibinfo{author}{\bibfnamefont{S.}~\bibnamefont{Sainis}},
  \bibinfo{author}{\bibfnamefont{T.}~\bibnamefont{Bergeman}}, \bibnamefont{and}
  \bibinfo{author}{\bibfnamefont{D.}~\bibnamefont{DeMille}},
  \bibinfo{journal}{Phys. Rev. Lett.} \textbf{\bibinfo{volume}{94}},
  \bibinfo{pages}{203001} (\bibinfo{year}{2005}).

\bibitem[{\citenamefont{Ni et~al.}(2008)\citenamefont{Ni, Ospelkaus, {de
  Miranda}, Pe'er, Neyenhuis, Zirbel, Kotochigova, Julienne, Jin, and
  Ye}}]{Ni:KRb:2008}
\bibinfo{author}{\bibfnamefont{K.-K.} \bibnamefont{Ni}},
  \bibinfo{author}{\bibfnamefont{S.}~\bibnamefont{Ospelkaus}},
  \bibinfo{author}{\bibfnamefont{M.~H.~G.} \bibnamefont{{de Miranda}}},
  \bibinfo{author}{\bibfnamefont{A.}~\bibnamefont{Pe'er}},
  \bibinfo{author}{\bibfnamefont{B.}~\bibnamefont{Neyenhuis}},
  \bibinfo{author}{\bibfnamefont{J.~J.} \bibnamefont{Zirbel}},
  \bibinfo{author}{\bibfnamefont{S.}~\bibnamefont{Kotochigova}},
  \bibinfo{author}{\bibfnamefont{P.~S.} \bibnamefont{Julienne}},
  \bibinfo{author}{\bibfnamefont{D.~S.} \bibnamefont{Jin}}, \bibnamefont{and}
  \bibinfo{author}{\bibfnamefont{J.}~\bibnamefont{Ye}},
  \bibinfo{journal}{Science} \textbf{\bibinfo{volume}{322}},
  \bibinfo{pages}{231} (\bibinfo{year}{2008}).

\bibitem[{\citenamefont{Lang et~al.}(2008)\citenamefont{Lang, Winkler, Strauss,
  Grimm, and Hecker~Denschlag}}]{Lang:ground:2008}
\bibinfo{author}{\bibfnamefont{F.}~\bibnamefont{Lang}},
  \bibinfo{author}{\bibfnamefont{K.}~\bibnamefont{Winkler}},
  \bibinfo{author}{\bibfnamefont{C.}~\bibnamefont{Strauss}},
  \bibinfo{author}{\bibfnamefont{R.}~\bibnamefont{Grimm}}, \bibnamefont{and}
  \bibinfo{author}{\bibfnamefont{J.}~\bibnamefont{Hecker~Denschlag}},
  \bibinfo{journal}{Phys. Rev. Lett.} \textbf{\bibinfo{volume}{101}},
  \bibinfo{pages}{133005} (\bibinfo{year}{2008}).

\bibitem[{\citenamefont{Danzl et~al.}(2008)\citenamefont{Danzl, Haller,
  Gustavsson, Mark, Hart, Bouloufa, Dulieu, Ritsch, and
  N\"agerl}}]{Danzl:v73:2008}
\bibinfo{author}{\bibfnamefont{J.~G.} \bibnamefont{Danzl}},
  \bibinfo{author}{\bibfnamefont{E.}~\bibnamefont{Haller}},
  \bibinfo{author}{\bibfnamefont{M.}~\bibnamefont{Gustavsson}},
  \bibinfo{author}{\bibfnamefont{M.~J.} \bibnamefont{Mark}},
  \bibinfo{author}{\bibfnamefont{R.}~\bibnamefont{Hart}},
  \bibinfo{author}{\bibfnamefont{N.}~\bibnamefont{Bouloufa}},
  \bibinfo{author}{\bibfnamefont{O.}~\bibnamefont{Dulieu}},
  \bibinfo{author}{\bibfnamefont{H.}~\bibnamefont{Ritsch}}, \bibnamefont{and}
  \bibinfo{author}{\bibfnamefont{H.-C.} \bibnamefont{N\"agerl}},
  \bibinfo{journal}{Science} \textbf{\bibinfo{volume}{321}},
  \bibinfo{pages}{1062} (\bibinfo{year}{2008}).

\bibitem[{\citenamefont{Viteau et~al.}(2008)\citenamefont{Viteau, Chotia,
  Allegrini, Bouloufa, Dulieu, Comparat, and Pillet}}]{Viteau:2008}
\bibinfo{author}{\bibfnamefont{M.}~\bibnamefont{Viteau}},
  \bibinfo{author}{\bibfnamefont{A.}~\bibnamefont{Chotia}},
  \bibinfo{author}{\bibfnamefont{M.}~\bibnamefont{Allegrini}},
  \bibinfo{author}{\bibfnamefont{N.}~\bibnamefont{Bouloufa}},
  \bibinfo{author}{\bibfnamefont{O.}~\bibnamefont{Dulieu}},
  \bibinfo{author}{\bibfnamefont{D.}~\bibnamefont{Comparat}}, \bibnamefont{and}
  \bibinfo{author}{\bibfnamefont{P.}~\bibnamefont{Pillet}},
  \bibinfo{journal}{Science} \textbf{\bibinfo{volume}{321}},
  \bibinfo{pages}{232} (\bibinfo{year}{2008}).

\bibitem[{\citenamefont{Deiglmayr et~al.}(2008)\citenamefont{Deiglmayr,
  Grochola, Repp, M\"ortlbauer, Gl\"uck, Lange, Dulieu, Wester, and
  Weidem\"uller}}]{Deiglmayr:2008}
\bibinfo{author}{\bibfnamefont{J.}~\bibnamefont{Deiglmayr}},
  \bibinfo{author}{\bibfnamefont{A.}~\bibnamefont{Grochola}},
  \bibinfo{author}{\bibfnamefont{M.}~\bibnamefont{Repp}},
  \bibinfo{author}{\bibfnamefont{K.}~\bibnamefont{M\"ortlbauer}},
  \bibinfo{author}{\bibfnamefont{C.}~\bibnamefont{Gl\"uck}},
  \bibinfo{author}{\bibfnamefont{J.}~\bibnamefont{Lange}},
  \bibinfo{author}{\bibfnamefont{O.}~\bibnamefont{Dulieu}},
  \bibinfo{author}{\bibfnamefont{R.}~\bibnamefont{Wester}}, \bibnamefont{and}
  \bibinfo{author}{\bibfnamefont{M.}~\bibnamefont{Weidem\"uller}},
  \bibinfo{journal}{Phys. Rev. Lett.} \textbf{\bibinfo{volume}{101}},
  \bibinfo{pages}{133004} (\bibinfo{year}{2008}).

\bibitem[{\citenamefont{Mark et~al.}(2009)\citenamefont{Mark, Danzl, Haller,
  Gustavsson, Bouloufa, Dulieu, Salami, Bergeman, Ritsch, Hart
  et~al.}}]{Mark:2009}
\bibinfo{author}{\bibfnamefont{M.~J.} \bibnamefont{Mark}},
  \bibinfo{author}{\bibfnamefont{J.~G.} \bibnamefont{Danzl}},
  \bibinfo{author}{\bibfnamefont{E.}~\bibnamefont{Haller}},
  \bibinfo{author}{\bibfnamefont{M.}~\bibnamefont{Gustavsson}},
  \bibinfo{author}{\bibfnamefont{N.}~\bibnamefont{Bouloufa}},
  \bibinfo{author}{\bibfnamefont{O.}~\bibnamefont{Dulieu}},
  \bibinfo{author}{\bibfnamefont{H.}~\bibnamefont{Salami}},
  \bibinfo{author}{\bibfnamefont{T.}~\bibnamefont{Bergeman}},
  \bibinfo{author}{\bibfnamefont{H.}~\bibnamefont{Ritsch}},
  \bibinfo{author}{\bibfnamefont{R.}~\bibnamefont{Hart}}, \bibnamefont{et~al.},
  \bibinfo{journal}{Appl. Phys. B} \textbf{\bibinfo{volume}{95}},
  \bibinfo{pages}{219} (\bibinfo{year}{2009}).

\bibitem[{\citenamefont{Haimberger et~al.}(2009)\citenamefont{Haimberger,
  Kleinert, Zabawa, Wakin, and Bigelow}}]{Haimberger:2009}
\bibinfo{author}{\bibfnamefont{C.}~\bibnamefont{Haimberger}},
  \bibinfo{author}{\bibfnamefont{J.}~\bibnamefont{Kleinert}},
  \bibinfo{author}{\bibfnamefont{P.}~\bibnamefont{Zabawa}},
  \bibinfo{author}{\bibfnamefont{A.}~\bibnamefont{Wakin}}, \bibnamefont{and}
  \bibinfo{author}{\bibfnamefont{N.~P.} \bibnamefont{Bigelow}},
  \bibinfo{journal}{New. J. Phys.} \textbf{\bibinfo{volume}{11}},
  \bibinfo{pages}{055042} (\bibinfo{year}{2009}).

\bibitem[{\citenamefont{Danzl et~al.}(2010)\citenamefont{Danzl, Mark, Haller,
  Gustavsson, Hart, Aldegunde, Hutson, and N\"agerl}}]{Danzl:ground:2010}
\bibinfo{author}{\bibfnamefont{J.~G.} \bibnamefont{Danzl}},
  \bibinfo{author}{\bibfnamefont{M.~J.} \bibnamefont{Mark}},
  \bibinfo{author}{\bibfnamefont{E.}~\bibnamefont{Haller}},
  \bibinfo{author}{\bibfnamefont{M.}~\bibnamefont{Gustavsson}},
  \bibinfo{author}{\bibfnamefont{R.}~\bibnamefont{Hart}},
  \bibinfo{author}{\bibfnamefont{J.}~\bibnamefont{Aldegunde}},
  \bibinfo{author}{\bibfnamefont{J.~M.} \bibnamefont{Hutson}},
  \bibnamefont{and} \bibinfo{author}{\bibfnamefont{H.-C.}
  \bibnamefont{N\"agerl}}, \bibinfo{journal}{Nature Physics}
  \textbf{\bibinfo{volume}{6}}, \bibinfo{pages}{265} (\bibinfo{year}{2010}).

\bibitem[{\citenamefont{Carr et~al.}(2009)\citenamefont{Carr, {DeMille}, Krems,
  and Ye}}]{Carr:NJPintro:2009}
\bibinfo{author}{\bibfnamefont{L.~D.} \bibnamefont{Carr}},
  \bibinfo{author}{\bibfnamefont{D.}~\bibnamefont{{DeMille}}},
  \bibinfo{author}{\bibfnamefont{R.~V.} \bibnamefont{Krems}}, \bibnamefont{and}
  \bibinfo{author}{\bibfnamefont{J.}~\bibnamefont{Ye}}, \bibinfo{journal}{New
  J. Phys.} \textbf{\bibinfo{volume}{11}}, \bibinfo{pages}{055049}
  (\bibinfo{year}{2009}).

\bibitem[{\citenamefont{Ospelkaus et~al.}(2010)\citenamefont{Ospelkaus, Ni,
  Wang, de~Miranda, Neyenhuis, Qu\'{e}m\'{e}ner, Julienne, Bohn, Jin, and
  Ye}}]{Ospelkaus:react:2010}
\bibinfo{author}{\bibfnamefont{S.}~\bibnamefont{Ospelkaus}},
  \bibinfo{author}{\bibfnamefont{K.-K.} \bibnamefont{Ni}},
  \bibinfo{author}{\bibfnamefont{D.}~\bibnamefont{Wang}},
  \bibinfo{author}{\bibfnamefont{M.~H.~G.} \bibnamefont{de~Miranda}},
  \bibinfo{author}{\bibfnamefont{B.}~\bibnamefont{Neyenhuis}},
  \bibinfo{author}{\bibfnamefont{G.}~\bibnamefont{Qu\'{e}m\'{e}ner}},
  \bibinfo{author}{\bibfnamefont{P.~S.} \bibnamefont{Julienne}},
  \bibinfo{author}{\bibfnamefont{J.~L.} \bibnamefont{Bohn}},
  \bibinfo{author}{\bibfnamefont{D.~S.} \bibnamefont{Jin}}, \bibnamefont{and}
  \bibinfo{author}{\bibfnamefont{J.}~\bibnamefont{Ye}},
  \bibinfo{journal}{Science} \textbf{\bibinfo{volume}{327}},
  \bibinfo{pages}{853} (\bibinfo{year}{2010}).

\bibitem[{\citenamefont{Sold\'{a}n et~al.}(2003)\citenamefont{Sold\'{a}n,
  Cvita\v{s}, and Hutson}}]{Soldan:2003}
\bibinfo{author}{\bibfnamefont{P.}~\bibnamefont{Sold\'{a}n}},
  \bibinfo{author}{\bibfnamefont{M.~T.} \bibnamefont{Cvita\v{s}}},
  \bibnamefont{and} \bibinfo{author}{\bibfnamefont{J.~M.}
  \bibnamefont{Hutson}}, \bibinfo{journal}{Phys. Rev. A}
  \textbf{\bibinfo{volume}{67}}, \bibinfo{pages}{054702}
  (\bibinfo{year}{2003}).

\bibitem[{\citenamefont{Coxon and Melville}(2006)}]{Coxon:2006}
\bibinfo{author}{\bibfnamefont{J.~A.} \bibnamefont{Coxon}} \bibnamefont{and}
  \bibinfo{author}{\bibfnamefont{T.~C.} \bibnamefont{Melville}},
  \bibinfo{journal}{J. Mol. Spectrosc.} \textbf{\bibinfo{volume}{235}},
  \bibinfo{pages}{235} (\bibinfo{year}{2006}).

\bibitem[{\citenamefont{Fellows}(1991)}]{Fellows:1991}
\bibinfo{author}{\bibfnamefont{C.~E.} \bibnamefont{Fellows}},
  \bibinfo{journal}{J. Chem. Phys.} \textbf{\bibinfo{volume}{94}},
  \bibinfo{pages}{5855} (\bibinfo{year}{1991}).

\bibitem[{\citenamefont{Tiemann et~al.}(2009)\citenamefont{Tiemann, Kn\"ockel,
  Kowalczyk, Jastrzebski, Pashov, Salami, and Ross}}]{Tiemann:2009}
\bibinfo{author}{\bibfnamefont{E.}~\bibnamefont{Tiemann}},
  \bibinfo{author}{\bibfnamefont{H.}~\bibnamefont{Kn\"ockel}},
  \bibinfo{author}{\bibfnamefont{P.}~\bibnamefont{Kowalczyk}},
  \bibinfo{author}{\bibfnamefont{W.}~\bibnamefont{Jastrzebski}},
  \bibinfo{author}{\bibfnamefont{A.}~\bibnamefont{Pashov}},
  \bibinfo{author}{\bibfnamefont{H.}~\bibnamefont{Salami}}, \bibnamefont{and}
  \bibinfo{author}{\bibfnamefont{A.~J.} \bibnamefont{Ross}},
  \bibinfo{journal}{Phys. Rev. A} \textbf{\bibinfo{volume}{79}},
  \bibinfo{pages}{042716} (\bibinfo{year}{2009}).

\bibitem[{\citenamefont{Grochola et~al.}(2009)\citenamefont{Grochola, Pashov,
  Deiglmayr, Repp, Tiemann, Wester, and Weidem\"{u}ller}}]{Grochola:2009}
\bibinfo{author}{\bibfnamefont{A.}~\bibnamefont{Grochola}},
  \bibinfo{author}{\bibfnamefont{A.}~\bibnamefont{Pashov}},
  \bibinfo{author}{\bibfnamefont{J.}~\bibnamefont{Deiglmayr}},
  \bibinfo{author}{\bibfnamefont{M.}~\bibnamefont{Repp}},
  \bibinfo{author}{\bibfnamefont{E.}~\bibnamefont{Tiemann}},
  \bibinfo{author}{\bibfnamefont{R.}~\bibnamefont{Wester}}, \bibnamefont{and}
  \bibinfo{author}{\bibfnamefont{M.}~\bibnamefont{Weidem\"{u}ller}},
  \bibinfo{journal}{J. Chem. Phys.} \textbf{\bibinfo{volume}{131}},
  \bibinfo{pages}{054304} (\bibinfo{year}{2009}).

\bibitem[{\citenamefont{Jones et~al.}(1996)\citenamefont{Jones, Maleki, Bize,
  Lett, Williams, Richling, Kn\"ockel, Tiemann, Wang, Gould
  et~al.}}]{Jones:1996}
\bibinfo{author}{\bibfnamefont{K.~M.} \bibnamefont{Jones}},
  \bibinfo{author}{\bibfnamefont{S.}~\bibnamefont{Maleki}},
  \bibinfo{author}{\bibfnamefont{S.}~\bibnamefont{Bize}},
  \bibinfo{author}{\bibfnamefont{P.~D.} \bibnamefont{Lett}},
  \bibinfo{author}{\bibfnamefont{C.~J.} \bibnamefont{Williams}},
  \bibinfo{author}{\bibfnamefont{H.}~\bibnamefont{Richling}},
  \bibinfo{author}{\bibfnamefont{H.}~\bibnamefont{Kn\"ockel}},
  \bibinfo{author}{\bibfnamefont{E.}~\bibnamefont{Tiemann}},
  \bibinfo{author}{\bibfnamefont{H.}~\bibnamefont{Wang}},
  \bibinfo{author}{\bibfnamefont{P.~L.} \bibnamefont{Gould}},
  \bibnamefont{et~al.}, \bibinfo{journal}{Phys. Rev. A}
  \textbf{\bibinfo{volume}{54}}, \bibinfo{pages}{R1006} (\bibinfo{year}{1996}).

\bibitem[{\citenamefont{Gerdes et~al.}(2008)\citenamefont{Gerdes, Hobein,
  Kn\"ockel, and Tiemann}}]{Gerdes:2008}
\bibinfo{author}{\bibfnamefont{A.}~\bibnamefont{Gerdes}},
  \bibinfo{author}{\bibfnamefont{M.}~\bibnamefont{Hobein}},
  \bibinfo{author}{\bibfnamefont{H.}~\bibnamefont{Kn\"ockel}},
  \bibnamefont{and} \bibinfo{author}{\bibfnamefont{E.}~\bibnamefont{Tiemann}},
  \bibinfo{journal}{Eur. Phys. J. D} \textbf{\bibinfo{volume}{49}},
  \bibinfo{pages}{67} (\bibinfo{year}{2008}).

\bibitem[{\citenamefont{Pashov et~al.}(2005)\citenamefont{Pashov, Docenko,
  Tamanis, Ferber, Kn\"ockel, and Tiemann}}]{Pashov:NaRb:2005}
\bibinfo{author}{\bibfnamefont{A.}~\bibnamefont{Pashov}},
  \bibinfo{author}{\bibfnamefont{O.}~\bibnamefont{Docenko}},
  \bibinfo{author}{\bibfnamefont{M.}~\bibnamefont{Tamanis}},
  \bibinfo{author}{\bibfnamefont{R.}~\bibnamefont{Ferber}},
  \bibinfo{author}{\bibfnamefont{H.}~\bibnamefont{Kn\"ockel}},
  \bibnamefont{and} \bibinfo{author}{\bibfnamefont{E.}~\bibnamefont{Tiemann}},
  \bibinfo{journal}{Phys. Rev. A} \textbf{\bibinfo{volume}{72}},
  \bibinfo{pages}{062505} (\bibinfo{year}{2005}).

\bibitem[{\citenamefont{Docenko et~al.}(2006)\citenamefont{Docenko, Tamanis,
  Zaharova, Ferber, Pashov, Kn\"ockel, and Tiemann}}]{Docenko:2006}
\bibinfo{author}{\bibfnamefont{O.}~\bibnamefont{Docenko}},
  \bibinfo{author}{\bibfnamefont{M.}~\bibnamefont{Tamanis}},
  \bibinfo{author}{\bibfnamefont{J.}~\bibnamefont{Zaharova}},
  \bibinfo{author}{\bibfnamefont{R.}~\bibnamefont{Ferber}},
  \bibinfo{author}{\bibfnamefont{A.}~\bibnamefont{Pashov}},
  \bibinfo{author}{\bibfnamefont{H.}~\bibnamefont{Kn\"ockel}},
  \bibnamefont{and} \bibinfo{author}{\bibfnamefont{E.}~\bibnamefont{Tiemann}},
  \bibinfo{journal}{J. Phys. B} \textbf{\bibinfo{volume}{39}},
  \bibinfo{pages}{S929} (\bibinfo{year}{2006}).

\bibitem[{\citenamefont{Pashov et~al.}(2008)\citenamefont{Pashov, Popov,
  Kn\"ockel, and Tiemann}}]{Pashov:2008}
\bibinfo{author}{\bibfnamefont{A.}~\bibnamefont{Pashov}},
  \bibinfo{author}{\bibfnamefont{P.}~\bibnamefont{Popov}},
  \bibinfo{author}{\bibfnamefont{H.}~\bibnamefont{Kn\"ockel}},
  \bibnamefont{and} \bibinfo{author}{\bibfnamefont{E.}~\bibnamefont{Tiemann}},
  \bibinfo{journal}{Eur. Phys. J. D} \textbf{\bibinfo{volume}{46}},
  \bibinfo{pages}{241} (\bibinfo{year}{2008}).

\bibitem[{\citenamefont{Pashov et~al.}(2007)\citenamefont{Pashov, Docenko,
  Tamanis, Ferber, Kn\"ockel, and Tiemann}}]{Pashov:2007}
\bibinfo{author}{\bibfnamefont{A.}~\bibnamefont{Pashov}},
  \bibinfo{author}{\bibfnamefont{O.}~\bibnamefont{Docenko}},
  \bibinfo{author}{\bibfnamefont{M.}~\bibnamefont{Tamanis}},
  \bibinfo{author}{\bibfnamefont{R.}~\bibnamefont{Ferber}},
  \bibinfo{author}{\bibfnamefont{H.}~\bibnamefont{Kn\"ockel}},
  \bibnamefont{and} \bibinfo{author}{\bibfnamefont{E.}~\bibnamefont{Tiemann}},
  \bibinfo{journal}{Phys. Rev. A} \textbf{\bibinfo{volume}{76}},
  \bibinfo{pages}{022511} (\bibinfo{year}{2007}).

\bibitem[{\citenamefont{Ferber et~al.}(2009)\citenamefont{Ferber, Klincare,
  Nikolayeva, Tamanis, Kn\"ockel, Tiemann, and Pashov}}]{Ferber:2009}
\bibinfo{author}{\bibfnamefont{R.}~\bibnamefont{Ferber}},
  \bibinfo{author}{\bibfnamefont{I.}~\bibnamefont{Klincare}},
  \bibinfo{author}{\bibfnamefont{O.}~\bibnamefont{Nikolayeva}},
  \bibinfo{author}{\bibfnamefont{M.}~\bibnamefont{Tamanis}},
  \bibinfo{author}{\bibfnamefont{H.}~\bibnamefont{Kn\"ockel}},
  \bibinfo{author}{\bibfnamefont{E.}~\bibnamefont{Tiemann}}, \bibnamefont{and}
  \bibinfo{author}{\bibfnamefont{A.}~\bibnamefont{Pashov}},
  \bibinfo{journal}{Phys. Rev. A} \textbf{\bibinfo{volume}{80}},
  \bibinfo{pages}{062501} (\bibinfo{year}{2009}).

\bibitem[{\citenamefont{Seto et~al.}(2000)\citenamefont{Seto, Le~Roy, Verg\`es,
  and Amiot}}]{Seto:2000}
\bibinfo{author}{\bibfnamefont{J.~Y.} \bibnamefont{Seto}},
  \bibinfo{author}{\bibfnamefont{R.~J.} \bibnamefont{Le~Roy}},
  \bibinfo{author}{\bibfnamefont{J.}~\bibnamefont{Verg\`es}}, \bibnamefont{and}
  \bibinfo{author}{\bibfnamefont{C.}~\bibnamefont{Amiot}}, \bibinfo{journal}{J.
  Chem. Phys} \textbf{\bibinfo{volume}{113}}, \bibinfo{pages}{3067}
  (\bibinfo{year}{2000}).

\bibitem[{\citenamefont{Fellows et~al.}(1999)\citenamefont{Fellows, Gutterres,
  Campos, Verg\`es, and Amiot}}]{Fellows:1999}
\bibinfo{author}{\bibfnamefont{C.~E.} \bibnamefont{Fellows}},
  \bibinfo{author}{\bibfnamefont{R.~F.} \bibnamefont{Gutterres}},
  \bibinfo{author}{\bibfnamefont{A.~P.~C.} \bibnamefont{Campos}},
  \bibinfo{author}{\bibfnamefont{J.}~\bibnamefont{Verg\`es}}, \bibnamefont{and}
  \bibinfo{author}{\bibfnamefont{C.}~\bibnamefont{Amiot}}, \bibinfo{journal}{J.
  Mol. Spectrosc.} \textbf{\bibinfo{volume}{197}}, \bibinfo{pages}{19}
  (\bibinfo{year}{1999}).

\bibitem[{\citenamefont{Amiot and Dulieu}(2002)}]{Amiot:2002}
\bibinfo{author}{\bibfnamefont{C.}~\bibnamefont{Amiot}} \bibnamefont{and}
  \bibinfo{author}{\bibfnamefont{O.}~\bibnamefont{Dulieu}},
  \bibinfo{journal}{J. Chem. Phys.} \textbf{\bibinfo{volume}{117}},
  \bibinfo{pages}{5155} (\bibinfo{year}{2002}).

\bibitem[{\citenamefont{Wu}(1976)}]{Wu:1976}
\bibinfo{author}{\bibfnamefont{C.~H.} \bibnamefont{Wu}}, \bibinfo{journal}{J.
  Chem. Phys.} \textbf{\bibinfo{volume}{65}}, \bibinfo{pages}{3181}
  (\bibinfo{year}{1976}).

\bibitem[{\citenamefont{Werner et~al.}(2006)\citenamefont{Werner, Knowles,
  Lindh, {Sch\"{u}tz} et~al.}}]{MOLPRO_brief:2006}
\bibinfo{author}{\bibfnamefont{H.-J.} \bibnamefont{Werner}},
  \bibinfo{author}{\bibfnamefont{P.~J.} \bibnamefont{Knowles}},
  \bibinfo{author}{\bibfnamefont{R.}~\bibnamefont{Lindh}},
  \bibinfo{author}{\bibfnamefont{M.}~\bibnamefont{{Sch\"{u}tz}}},
  \bibnamefont{et~al.}, \emph{\bibinfo{title}{{\sc MOLPRO}, version 2006.1: A
  package of ab initio programs}} (\bibinfo{year}{2006}), \bibinfo{note}{see
  http://www.molpro.net}.

\bibitem[{\citenamefont{Fuentealba et~al.}(1982)\citenamefont{Fuentealba,
  Preuss, Stoll, and {von Szentpaly}}}]{Fuentealba:CPP1:1982}
\bibinfo{author}{\bibfnamefont{P.}~\bibnamefont{Fuentealba}},
  \bibinfo{author}{\bibfnamefont{H.}~\bibnamefont{Preuss}},
  \bibinfo{author}{\bibfnamefont{H.}~\bibnamefont{Stoll}}, \bibnamefont{and}
  \bibinfo{author}{\bibfnamefont{L.}~\bibnamefont{{von Szentpaly}}},
  \bibinfo{journal}{Chem. Phys. Lett.} \textbf{\bibinfo{volume}{89}},
  \bibinfo{pages}{418} (\bibinfo{year}{1982}).

\bibitem[{\citenamefont{Fuentealba et~al.}(1983)\citenamefont{Fuentealba,
  Stoll, {von Szentpaly}, Schwerdtfeger, and Preuss}}]{Fuentealba:CPP1:1983}
\bibinfo{author}{\bibfnamefont{P.}~\bibnamefont{Fuentealba}},
  \bibinfo{author}{\bibfnamefont{H.}~\bibnamefont{Stoll}},
  \bibinfo{author}{\bibfnamefont{L.}~\bibnamefont{{von Szentpaly}}},
  \bibinfo{author}{\bibfnamefont{P.}~\bibnamefont{Schwerdtfeger}},
  \bibnamefont{and} \bibinfo{author}{\bibfnamefont{H.}~\bibnamefont{Preuss}},
  \bibinfo{journal}{J. Phys. B} \textbf{\bibinfo{volume}{16}},
  \bibinfo{pages}{L323} (\bibinfo{year}{1983}).

\bibitem[{\citenamefont{M\"uller et~al.}(1984)\citenamefont{M\"uller, Flesch,
  and Meyer}}]{Muller:CPP2:1984}
\bibinfo{author}{\bibfnamefont{W.}~\bibnamefont{M\"uller}},
  \bibinfo{author}{\bibfnamefont{J.}~\bibnamefont{Flesch}}, \bibnamefont{and}
  \bibinfo{author}{\bibfnamefont{W.}~\bibnamefont{Meyer}}, \bibinfo{journal}{J.
  Chem. Phys.} \textbf{\bibinfo{volume}{80}}, \bibinfo{pages}{3297}
  (\bibinfo{year}{1984}).

\bibitem[{alk()}]{alkali-basis:CPP:2010}
\bibinfo{note}{$s$ functions with exponents 0.010159 for Li, 0.009202 for Na,
  0.009433 for K, 0.007182 for Rb, 0.009778 for Cs, $p$ functions with
  exponents 0.007058 for Li, 0.005306 for Na, 0.004358 for K, 0.004459 for Rb,
  0.004186 for Cs, $d$ functions with exponents 0.39 and 0.13 for Li, 0.3 and
  0.1 for Na, 0.27 and 0.09 for K, 0.21 and 0.07 for both Rb and Cs and $f$
  functions with exponents 0.13 for Li, 0.1 for Na, 0.09 for K, 0.07 for both
  Rb and Cs.}

\bibitem[{tri()}]{trimer-EPAPS}
\bibinfo{note}{See EPAPS Document No. [number will be inserted by publisher]
  for tabulations of equilibrium geometries and energies for $^2A_1$ and
  $^2B_2$ states. For more information on EPAPS, see
  http://www.aip.org/pubservs/epaps.html.}

\bibitem[{\citenamefont{Tscherbul et~al.}(2008)\citenamefont{Tscherbul,
  Barinovs, K{\l}os, and Krems}}]{Tscherbul:Rb2Cs:2008}
\bibinfo{author}{\bibfnamefont{T.~V.} \bibnamefont{Tscherbul}},
  \bibinfo{author}{\bibfnamefont{{\v G}.}~\bibnamefont{Barinovs}},
  \bibinfo{author}{\bibfnamefont{J.}~\bibnamefont{K{\l}os}}, \bibnamefont{and}
  \bibinfo{author}{\bibfnamefont{R.~V.} \bibnamefont{Krems}},
  \bibinfo{journal}{Phys. Rev. A} \textbf{\bibinfo{volume}{78}},
  \bibinfo{pages}{022705} (\bibinfo{year}{2008}).

\bibitem[{\citenamefont{Sold\'{a}n et~al.}(2002)\citenamefont{Sold\'{a}n,
  Cvita\v{s}, Hutson, Honvault, and Launay}}]{Soldan:2002}
\bibinfo{author}{\bibfnamefont{P.}~\bibnamefont{Sold\'{a}n}},
  \bibinfo{author}{\bibfnamefont{M.~T.} \bibnamefont{Cvita\v{s}}},
  \bibinfo{author}{\bibfnamefont{J.~M.} \bibnamefont{Hutson}},
  \bibinfo{author}{\bibfnamefont{P.}~\bibnamefont{Honvault}}, \bibnamefont{and}
  \bibinfo{author}{\bibfnamefont{J.~M.} \bibnamefont{Launay}},
  \bibinfo{journal}{Phys. Rev. Lett.} \textbf{\bibinfo{volume}{89}},
  \bibinfo{pages}{153201} (\bibinfo{year}{2002}).

\bibitem[{\citenamefont{Cvita\v{s}
  et~al.}(2005{\natexlab{a}})\citenamefont{Cvita\v{s}, Sold\'{a}n, Hutson,
  Honvault, and Launay}}]{Cvitas:bosefermi:2005}
\bibinfo{author}{\bibfnamefont{M.~T.} \bibnamefont{Cvita\v{s}}},
  \bibinfo{author}{\bibfnamefont{P.}~\bibnamefont{Sold\'{a}n}},
  \bibinfo{author}{\bibfnamefont{J.~M.} \bibnamefont{Hutson}},
  \bibinfo{author}{\bibfnamefont{P.}~\bibnamefont{Honvault}}, \bibnamefont{and}
  \bibinfo{author}{\bibfnamefont{J.~M.} \bibnamefont{Launay}},
  \bibinfo{journal}{Phys. Rev. Lett.} \textbf{\bibinfo{volume}{94}},
  \bibinfo{pages}{033201} (\bibinfo{year}{2005}{\natexlab{a}}).

\bibitem[{\citenamefont{Cvita\v{s}
  et~al.}(2005{\natexlab{b}})\citenamefont{Cvita\v{s}, Sold\'{a}n, Hutson,
  Honvault, and Launay}}]{Cvitas:hetero:2005}
\bibinfo{author}{\bibfnamefont{M.~T.} \bibnamefont{Cvita\v{s}}},
  \bibinfo{author}{\bibfnamefont{P.}~\bibnamefont{Sold\'{a}n}},
  \bibinfo{author}{\bibfnamefont{J.~M.} \bibnamefont{Hutson}},
  \bibinfo{author}{\bibfnamefont{P.}~\bibnamefont{Honvault}}, \bibnamefont{and}
  \bibinfo{author}{\bibfnamefont{J.~M.} \bibnamefont{Launay}},
  \bibinfo{journal}{Phys. Rev. Lett.} \textbf{\bibinfo{volume}{94}},
  \bibinfo{pages}{200402} (\bibinfo{year}{2005}{\natexlab{b}}).

\bibitem[{\citenamefont{Qu\'{e}m\'{e}ner
  et~al.}(2005)\citenamefont{Qu\'{e}m\'{e}ner, Honvault, Launay, Sold\'{a}n,
  Potter, and Hutson}}]{Quemener:2005}
\bibinfo{author}{\bibfnamefont{G.}~\bibnamefont{Qu\'{e}m\'{e}ner}},
  \bibinfo{author}{\bibfnamefont{P.}~\bibnamefont{Honvault}},
  \bibinfo{author}{\bibfnamefont{J.~M.} \bibnamefont{Launay}},
  \bibinfo{author}{\bibfnamefont{P.}~\bibnamefont{Sold\'{a}n}},
  \bibinfo{author}{\bibfnamefont{D.~E.} \bibnamefont{Potter}},
  \bibnamefont{and} \bibinfo{author}{\bibfnamefont{J.~M.}
  \bibnamefont{Hutson}}, \bibinfo{journal}{Phys. Rev. A}
  \textbf{\bibinfo{volume}{71}}, \bibinfo{pages}{032722}
  (\bibinfo{year}{2005}).

\bibitem[{\citenamefont{Cvita\v{s} et~al.}(2007)\citenamefont{Cvita\v{s},
  Sold\'{a}n, Hutson, Honvault, and Launay}}]{Cvitas:li3:2007}
\bibinfo{author}{\bibfnamefont{M.~T.} \bibnamefont{Cvita\v{s}}},
  \bibinfo{author}{\bibfnamefont{P.}~\bibnamefont{Sold\'{a}n}},
  \bibinfo{author}{\bibfnamefont{J.~M.} \bibnamefont{Hutson}},
  \bibinfo{author}{\bibfnamefont{P.}~\bibnamefont{Honvault}}, \bibnamefont{and}
  \bibinfo{author}{\bibfnamefont{J.~M.} \bibnamefont{Launay}},
  \bibinfo{journal}{J. Chem. Phys.} \textbf{\bibinfo{volume}{127}},
  \bibinfo{pages}{074302} (\bibinfo{year}{2007}).

\bibitem[{\citenamefont{Hutson and Sold\'{a}n}(2007)}]{Hutson:IRPC:2007}
\bibinfo{author}{\bibfnamefont{J.~M.} \bibnamefont{Hutson}} \bibnamefont{and}
  \bibinfo{author}{\bibfnamefont{P.}~\bibnamefont{Sold\'{a}n}},
  \bibinfo{journal}{Int. Rev. Phys. Chem.} \textbf{\bibinfo{volume}{26}},
  \bibinfo{pages}{1} (\bibinfo{year}{2007}).

\bibitem[{\citenamefont{Kr\"amer et~al.}(1999)\citenamefont{Kr\"amer, Keil,
  Suarez, Demtr\"oder, and Meyer}}]{Kramer:1999}
\bibinfo{author}{\bibfnamefont{H.-G.} \bibnamefont{Kr\"amer}},
  \bibinfo{author}{\bibfnamefont{M.}~\bibnamefont{Keil}},
  \bibinfo{author}{\bibfnamefont{C.~B.} \bibnamefont{Suarez}},
  \bibinfo{author}{\bibfnamefont{W.}~\bibnamefont{Demtr\"oder}},
  \bibnamefont{and} \bibinfo{author}{\bibfnamefont{W.}~\bibnamefont{Meyer}},
  \bibinfo{journal}{Chem. Phys. Lett.} \textbf{\bibinfo{volume}{299}},
  \bibinfo{pages}{212} (\bibinfo{year}{1999}).

\bibitem[{\citenamefont{Gu\'erout et~al.}(2009)\citenamefont{Gu\'erout,
  Sold\'an, Aymar, Deiglmayr, and Dulieu}}]{Guerout:2009}
\bibinfo{author}{\bibfnamefont{R.}~\bibnamefont{Gu\'erout}},
  \bibinfo{author}{\bibfnamefont{P.}~\bibnamefont{Sold\'an}},
  \bibinfo{author}{\bibfnamefont{M.}~\bibnamefont{Aymar}},
  \bibinfo{author}{\bibfnamefont{J.}~\bibnamefont{Deiglmayr}},
  \bibnamefont{and} \bibinfo{author}{\bibfnamefont{O.}~\bibnamefont{Dulieu}},
  \bibinfo{journal}{Int. J. Quantum Chem.} \textbf{\bibinfo{volume}{109}},
  \bibinfo{pages}{3387} (\bibinfo{year}{2009}).

\bibitem[{\citenamefont{Pavolini and Spiegelmann}(1987)}]{Pavolini:1987}
\bibinfo{author}{\bibfnamefont{D.}~\bibnamefont{Pavolini}} \bibnamefont{and}
  \bibinfo{author}{\bibfnamefont{F.}~\bibnamefont{Spiegelmann}},
  \bibinfo{journal}{J. Chem. Phys.} \textbf{\bibinfo{volume}{87}},
  \bibinfo{pages}{2854} (\bibinfo{year}{1987}).

\end{thebibliography}

\end{document}